# AI-AI co-creation outperforms human pairs in creative tasks

Yingyue Luna Luan[1], Luning Sun[2], Yeun Joon Kim[2,3]*, Jindong Wang[4], Xing Xie[5]

[1]UQ Business School, The University of Queensland, Brisbane, QLD 4067, Australia.

[2]Cambridge Judge Business School, University of Cambridge, Cambridge CB2 1AG, United Kingdom.

[3]Institute of Metabolic Science, School of Clinical Medicine, University of Cambridge, Cambridge CB2 0QQ, United Kingdom.

[4]Department of Data Science, William & Mary, Williamsburg, Virginia 23185, United States of America.

[5]Microsoft Research Asia, Beijing 100080, China.

*Corresponding author. Email: y.kim@jbs.cam.ac.uk

## Abstract

Prior research often finds that AI creativity is limited: single systems rarely outperform humans, and human-AI collaboration does not exceed human output. We argue these conclusions underestimate AI's potential because most studies do not allow iterative, multi-agent exchanges that mirror the social processes underpinning human creativity. We conducted an experiment comparing four conditions: (i) AI-AI co-creation with complementary generator-evaluator roles, (ii) AI-AI co-creation with identical roles, (iii) single-AI creation, and (iv) human-human co-creation. Across three open-ended tasks, 1,212 ideas were rated by trained judges on creativity, novelty, and usefulness. Both AI-AI co-creation conditions consistently outperformed single-AI creation and human pairs on creativity and novelty. Usefulness varied by task: complementary roles yielded the most useful solutions in the broadest and most socially complex task, suggesting role differentiation is advantageous when problems require both imaginative ideation and practical refinement. Human pairs performed worst, consistent with production losses in group creativity. These findings indicate that structured, iterative multi-agent AI co-creation can exceed single-AI and human-human ideation.

***Keywords***: AI-AI co-creation; creativity; multi-agent

Creativity does not emerge *ex nihilo*. It has long been understood as the outcome of social processes in which multiple agents—typically humans—interact to produce something novel and useful. Yet, in the era of artificial intelligence (AI), an important question arises: can multiple AI systems collaborate creatively in ways comparable to, or even surpassing, human-human co-creation? Unfortunately, extant literature does not offer a positive answer. Recent studies show that a single AI system's creativity does not exceed that of humans[1], nor does human-AI co-creation outperform human creativity[2,3]. We argue, however, that these studies do not fully leverage AI's creative potential. In studies comparing a single AI with humans, AIs were not permitted to refine or iteratively develop their ideas; instead, they were asked to generate an idea once, and their creative processes were immediately terminated. In studies comparing human-AI pairs with individual humans, researchers have shown that the lack of improvement in joint creativity largely stemmed from human biases (e.g., automation bias) and/or from human partners' difficulty in managing creative processes (e.g., insufficient engagement in joint idea refinement)[3]. Under these constraints, AI systems cannot demonstrate the generative and developmental aspects of their creative capacity.

Our research proposes that AI's creative potential is best realized when multiple AI agents interact across multiple rounds without human biases or interventions. We conceptualize AI-AI co-creation as an evolutionary process in which several AI agents assume the roles of idea generator and evaluator, jointly producing creative outcomes through iterative interactions. AI-AI co-creation offers several advantages, which can potentially surpass human creativity. First, AI systems can process information and generate ideas far more rapidly than humans, enabling AI-AI pairs to examine a substantially larger number of ideas. Second, multiple AI systems can collectively draw on extensive domains of data and knowledge, facilitating unusual combinations and unconventional ideas that may not arise in human ideation. Third—and most importantly—the dual roles of idea generation and evaluation enable AI agents not only to propose new ideas but also to engage in rigorous critique and refinement. As noted earlier, a central limitation of prior research is the absence of AI-driven evaluation. Creativity inherently relies on iterative cycles of idea generation and assessment, yet studies of AI creativity have largely prevented AI systems from critically examining or developing their own ideas. This omission is consequential, as the essence of creativity lies not merely in generating ideas but in continuously refining them[3].

To address this critical issue in the literature, we conducted an experimental study comparing creative performance across four conditions: (i) AI-AI co-creation with AI agents assigned complementary generator-evaluator roles, (ii) AI-AI co-creation with agents sharing identical roles (both as generator and evaluator), (iii) single-AI creation, and (iv) human-human co-creation. Across three open-ended creative tasks, all AI agents and human participants produced solutions of 50-100 words each. Trained human evaluators independently assessed these solutions on their creativity, novelty, and usefulness[4]. This work contributes to the growing field of AI creativity by directly comparing the creative processes and outcomes of human and AI systems. By studying the creative ideation in both human and AI contexts, we aim to deepen our understanding of various forms of creative collaboration while also exploring the computational capabilities and limitations of AI systems in creative tasks.

## Results

Across all conditions, human participants and AI agents generated solutions to three open-ended tasks frequently employed in creativity research. The experiment yielded 1,212 solutions, enabling comparison of creativity, novelty, and usefulness across the four conditions.

**Creativity**

Creativity scores differed significantly across conditions for all three tasks (task 1: $F(3, 398)=53.76$, $p<.001$; task 2: $F(3, 398)=33.53$, $p<.001$; task 3: $F(3, 397)=83.19$, $p<.001$; Figure 1). Across tasks, AI-AI conditions consistently outperformed human-human co-creation. Both forms of AI-AI co-creation yielded substantially higher creativity than human pairs in every task (all $p$s<.001). Single-AI creation exceeded human performance in tasks 1 and 2 (both $p$s<.001), but this advantage disappeared in task 3 ($p=.73$). Comparing AI conditions, AI-AI co-creation reliably matched or surpassed single-AI creation: both AI-AI conditions produced more creative ideas than single AI in tasks 1 and 3 (both $p$s<.01), whereas all three AI conditions performed similarly in task 2 ($p>.15$). Differences between the two AI-AI conditions were minimal in tasks 1 and 2 (both $p$s>.10), but the complementary-role condition outperformed the identical-role condition in task 3 ($p<.01$). Taken together, these results show that AI-AI co-creation consistently generates more creative output across tasks, with role differentiation providing additional benefits in the more socially complex task (i.e., task 3).

**Novelty**

Novelty scores differed significantly across conditions for all three tasks (task 1: $F(3, 398)=150.56$, $p<.001$; task 2: $F(3, 398)=92.66$, $p<.001$; task 3: $F(3, 398)=91.48$, $p<.001$; Figure 2). Across tasks, both forms of AI-AI co-creation produced the most novel ideas: in all three tasks, AI-AI co-creation with complementary roles and with identical roles yielded higher novelty than both human-human co-creation and single-AI creation ($p$s<.001). Single-AI-generated ideas were more novel than those by human pairs in tasks 1 and 2 (*p*s$<.01$), but this advantage disappeared in task 3 ($p=1.00$). Comparisons between the two AI-AI conditions revealed no significant differences in all tasks (*p*s$>.15$). These results indicate that AI-AI co-creation robustly enhances the novelty of generated ideas beyond both single-AI creation and human-human co-creation, whereas the novelty advantage of a single AI over human pairs is task-dependent.

**Usefulness**

Usefulness scores differed significantly across conditions for tasks 1 and 3 but only marginally for task 2 (task 1: $F(3, 398)=6.20$, $p<.001$; task 2: $F(3, 398)=2.46$, $p=.062$; task 3: $F(3, 398)=27.32$, $p<.001$; Figure 3). In task 1, ideas generated by AI-AI co-creation with complementary roles and by single-AI creation were more useful than those by human-human co-creation (*p*s$<.05$), whereas AI-AI co-creation with identical roles did not differ from human-human co-creation ($p=.29$). In task 2, only the single-AI condition produced more useful ideas than human-human co-creation ($p<.05$); both AI-AI conditions were comparable to human pairs and to single AI (*p*s$>0.31$). In task 3, all three AI conditions yielded more useful ideas than human-human co-creation (*p*s$<.001$). Moreover, AI-AI co-creation with complementary roles produced the most useful solutions, outperforming both AI-AI co-creation with identical roles and single-AI creation ($p<.001$), whereas the latter two did not differ from each other ($p=1.00$). Overall, although the pattern of results for usefulness is more mixed, AI-AI co-creation with complementary roles generally exhibited the strongest performance.

## Discussion

This study provides an exploratory comparison of creative ideation among different conditions of AI-AI co-creation, single-AI creation, and human-human co-creation across multiple open-ended tasks. In terms of creativity and novelty, the results were strikingly consistent: AI-AI co-creation outperformed both single AI and human pairs in every task. Usefulness showed a more nuanced pattern. Collectively, these findings offer new insights into the capability of multi-agent AI systems in creative work.

**AI-AI co-creation outperforms human-human co-creation and single-AI creation**

Across the three tasks, both forms of AI-AI co-creation produced ideas that were substantially more creative and novel than those generated either through single-AI creation or human–human co-creation. These findings reshape our understanding of AI's impact on creativity. Prior literature's failure to detect AI's creative potential largely stems from experimental designs that constrained AI systems from engaging in dynamic, iterative exchanges of ideas. Notably, even when two AI agents were assigned the identical roles, their creative performance remained high. This pattern suggests that AI-AI synergy does not depend solely on role specialization; rather, it emerges from the iterative exchange and refinement of ideas.

**Role differentiation is not always necessary, but it can matter sometimes**

Although complementary generator-evaluator roles did not enhance creativity or novelty beyond identical-role AI-AI co-creation, this configuration produced the most useful ideas in our task 3 (the water-saving task). This task had broader societal implications and was more socially complex than the other two, as it required not only divergent ideation but also solutions that were broadly useful. With these requirements, dedicated roles may be advantageous—where the idea generator AI agent focuses on generating novel possibilities, and the evaluator AI agent helps balance novelty with practicality through iterative feedback. Future research should further explore contexts in which role differentiation in AI-AI co-creation is most beneficial for producing ideas that are both novel and useful.

**Human-human co-creation underperformed AI-AI co-creation**

Human pairs consistently performed worse than the two AI-AI conditions on creativity and novelty, and often on usefulness as well. Several mechanisms may account for this pattern. Human pairs may self-censor or prematurely converge on safer ideas; social dynamics such as politeness norms, avoidance of disagreement, turn-taking, or coordination demands may constrain divergent thinking; and humans cannot match AI systems in the speed or breadth of associative search. Indeed, the creativity literature has documented that human collaboration is susceptible to production losses, including groupthink and social loafing[5,6]. By contrast, AI-AI co-creation is not influenced by divergent motivations or emotions. AI pairs remain focused exclusively on task goals and avoid human-like interpersonal noise, while also benefiting from substantially greater cognitive capacity and rapid information retrieval. Consequently, these features contribute to their ability to generate more creative ideas.

It is important to note that our study focused on a limited set of open-ended ideation tasks and did not consider the wide range of creative abilities that humans possess, such as emotional intelligence and contextual understanding[7,8]. Many creative domains, such as product design, UI/UX, marketing communication, or even storytelling, rely heavily on emotional resonance and subjective human experience. These forms of creativity require sensitivity to affect, empathy, and tacit cultural understanding—capacities that current AI systems may not fully emulate. The absence of such emotional complexity in our tasks may partly explain the strong performance of AI-AI co-creation relative to human pairs. Future research should investigate whether AI-AI pairs can perform comparably in creative domains where emotional or experiential qualities are central to evaluating the output.

Overall, our findings highlight the creative capabilities of AI-AI co-creation. AI-AI co-creation may serve as a powerful engine for early-stage ideation, enabling rapid exploration of diverse concepts at a scale and pace unattainable for human teams. Role-optimized AI systems may be especially valuable in domains that require balancing imaginative idea generation with critical evaluation, such as product design, engineering, policy development, or strategic planning.

## References

1. de Rooij, A. & Biskjaer, M. M. Has AI Surpassed Humans in Creative Idea Generation? A Meta-Analysis. in *Proceedings of the 36th Annual Conference of the European Association of Cognitive Ergonomics (EACE)* 1–11 (2025). doi:10.31234/osf.io/9u2ke_v1.
2. Vaccaro, M., Almaatouq, A. & Malone, T. When combinations of humans and AI are useful: A systematic review and meta-analysis. *Nat Hum Behav* **8**, 2293–2303 (2024).
3. Luan, Y. L., Kim, Y. J. & Zhou, J. Augmented Learning for Joint Creativity in Human-GenAI Co-Creation. *Information Systems Research* https://doi.org/10.1287/isre.2024.0984 (2025) doi:10.1287/isre.2024.0984.
4. Harvey, S. & Berry, J. Toward a meta-theory of creativity forms: How novelty and usefulness shape creativity. *AMR* https://doi.org/10.5465/amr.2020.0110 (2022) doi:10.5465/amr.2020.0110.
5. Turner, M. E. & Pratkanis, A. R. Twenty-Five Years of Groupthink Theory and Research: Lessons from the Evaluation of a Theory. *Organizational Behavior and Human Decision Processes* **73**, 105–115 (1998).
6. Karau, S. J. & Williams, K. D. Social loafing: A meta-analytic review and theoretical integration. *Journal of Personality and Social Psychology* **65**, 681–706 (1993).
7. Guilford, J. P. Creativity: Yesterday, Today and Tomorrow. *The Journal of Creative Behavior* **1**, 3–14 (1967).
8. Amabile, T. M. The social psychology of creativity: A componential conceptualization. *Journal of Personality and Social Psychology* **45**, 357–376 (1983).
9. Shalley, C. E. Effects of productivity goals, creativity goals, and personal discretion on individual creativity. *Journal of Applied Psychology* **76**, 179–185 (1991).

10. Molnar, A. SMARTRIQS: A Simple Method Allowing Real-Time Respondent Interaction in Qualtrics Surveys. *Journal of Behavioral and Experimental Finance* **22**, 161–169 (2019).

11. Amabile, T. M. Social psychology of creativity: A consensual assessment technique. *Journal of Personality and Social Psychology* **43**, 997–1013 (1982).

12. Kaufman, J. C., Baer, J. & Cole, J. C. Expertise, Domains, and the Consensual Assessment Technique. *The Journal of Creative Behavior* **43**, 223–233 (2009).

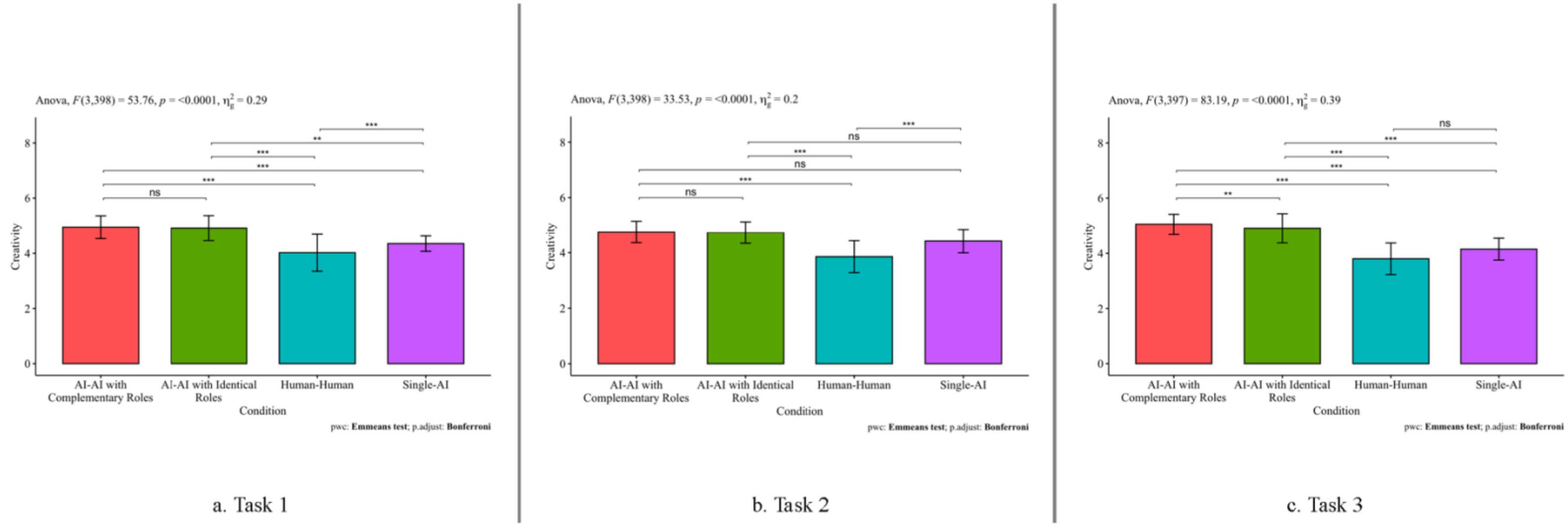

**Fig. 1.** ANOVA results and pairwise comparisons of creativity of ideas for each task

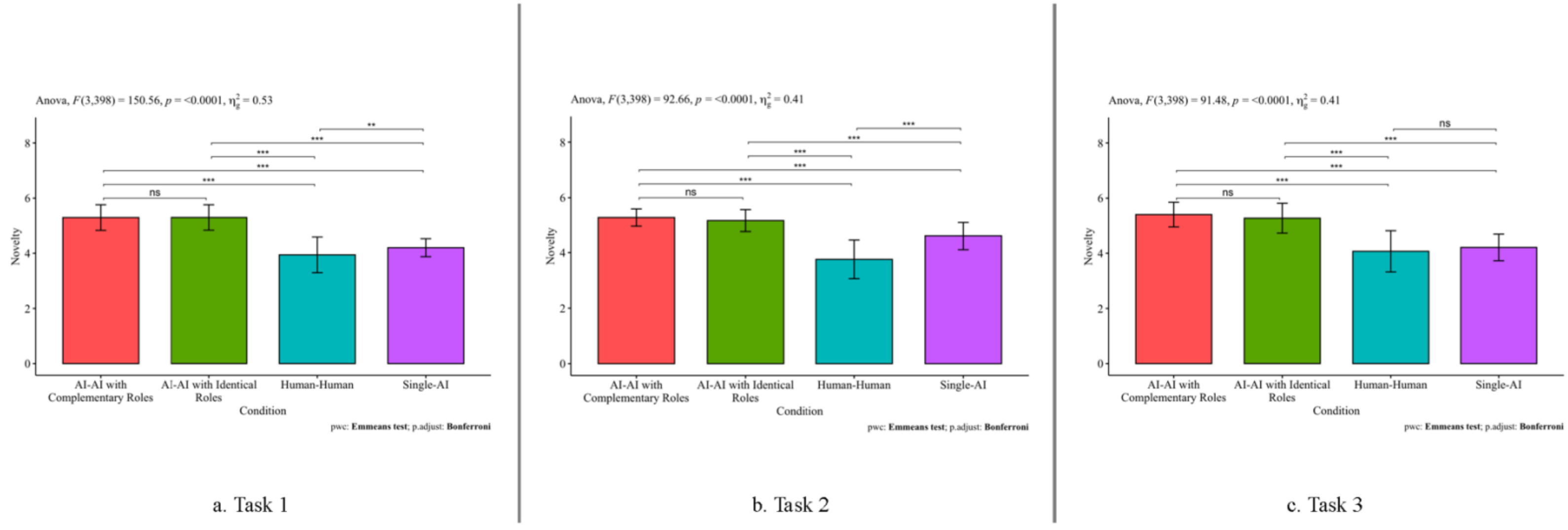

*** p<.01; *** p<.001; ns: not significant.*

**Fig. 2.** ANOVA results and pairwise comparisons of novelty of ideas for each task

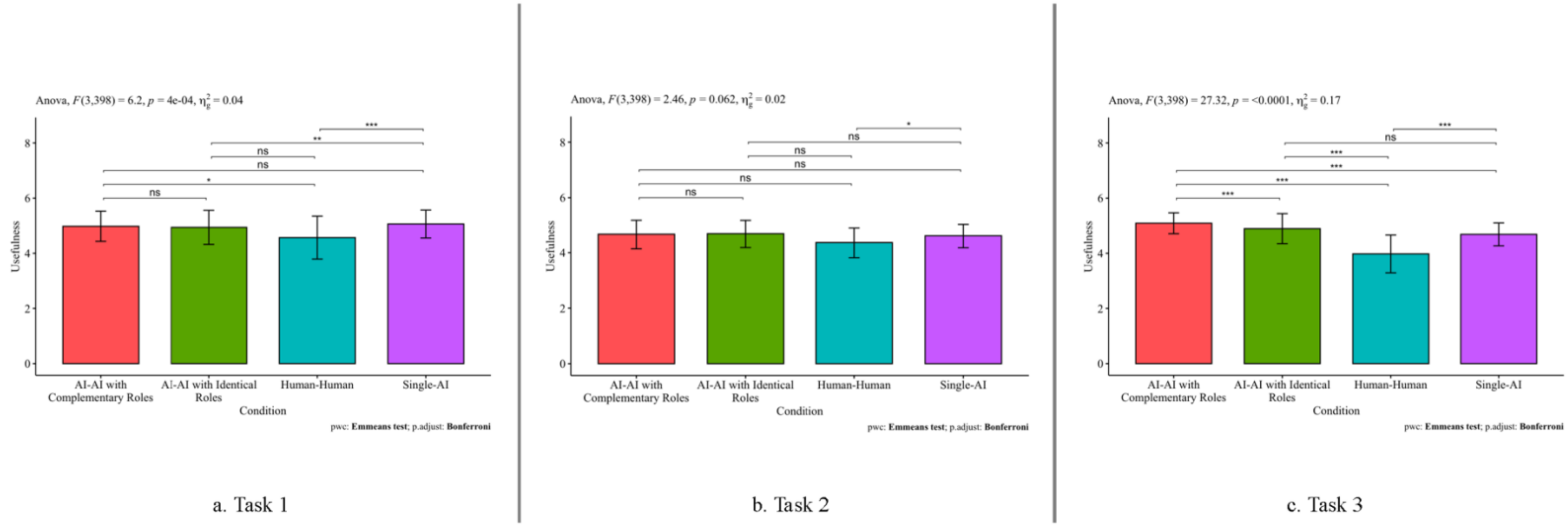

** p<.05; ** p<.01; *** p<.001; ns: not significant.*

**Fig. 3.** ANOVA results and pairwise comparisons of usefulness of ideas for each task

## Methods

### Study Design

We conducted an experimental study comparing the creative ideation process and outcomes between human-human and AI-AI co-creations. The study employed four conditions: i) AI-AI co-creation with agents taking complementary roles, ii) AI-AI co-creation with agents taking identical roles, iii) single-AI creation, and iv) human-human co-creation. The study applied the four conditions across three creative tasks. The tasks were two open-ended business problems related to improving employee cafeteria food quality[9] and addressing disruptive employee parties[9], and one open-ended social problem related to generating creative ways for society to save water. The details of the tasks can be found in Appendix A of the Supplementary Information.

**Condition 1:** AI-AI co-creation with agents taking complementary roles. In the first condition, two AI agents collaborated, and each took on a distinct role. The collaboration was conducted using the GPT-4 language model. The first AI agent was instructed to be the idea generator agent and was responsible for always generating creative solutions to the given problem. It was set to a temperature of 1, allowing for more diverse and exploratory idea generation. The other AI agent was the idea evaluator agent and was tasked with assessing the ideas proposed by the generator agent. It provided feedback, identified strengths and weaknesses, and assigned a creativity score on a scale of 1-10. The evaluator agent was set to a temperature of 0 to encourage more focused and critical evaluations.

The co-creation between the two AI agents followed an iterative process. First, the idea generator AI agent was prompted with the task description and instructed to generate one creative solution of 50-100 words. The generated solution was then passed on to the idea evaluator AI agent, along with the task description. The evaluator AI agent assessed the solution and provided a creativity score from 1-10, along with feedback on how to improve the creativity of the solution. If the evaluator AI assigned a creativity score lower than 10, the generator AI agent received the feedback and was prompted to revise the solution based on the evaluator's input. This process was repeated until the evaluator AI agent assigned a creativity score of 10 to the revised solution. On average, 3.8 interactions were taken between the two AI agents for task 1, 3.97 interactions for task 2, and 3.91 interactions for task 3.

For each co-creation, the following data was collected: the full conversation history between the agents, all the proposed solutions and their corresponding creativity scores, the number of iterations required to achieve a creativity score of 10, and the final solution that received a score of 10. This iterative co-creation process allowed the AI agents to build upon each other's ideas, refine solutions based on feedback, and strive for highly creative outcomes. By assigning complementary roles to the agents, we aimed to simulate a structured and productive collaboration dynamic.

**Condition 2:** AI-AI co-creation with agents taking identical roles. The second scenario mirrored the first, albeit with both AI agents adopting identical roles. Both AI agents in this condition were set to a temperature of 0.5, striking a balance between exploratory idea generation and focused refinement. The agents had identical roles and capabilities, allowing for a collaborative dynamic where both agents contributed equally to the ideation process.

The collaboration between the two agents again followed an iterative process. To begin, the first AI agent (Agent 1) was prompted with the task description and instructed to generate one creative solution of 50-100 words. The generated solution was then passed on to the second AI agent (Agent 2), along with the task description. Agent 2 assessed the solution provided by Agent 1 and offered feedback on how to improve its creativity. Additionally, Agent 2 generated a revised solution based on Agent 1's initial idea and the identified areas for improvement. This revised solution was then passed back to Agent 1. Agent 1 reviewed the feedback and the revised solution provided by Agent 2. It then generated its own revised solution, building upon the ideas and improvements suggested by Agent 2. This process of iterative feedback and solution refinement continued until one of the agents determined that the solution had reached a creativity score of 10 out of 10. On average, 4.29 interactions were taken between the two AI agents for task 1, 4.68 interactions for task 2, and 4.26 interactions for task 3. We showed the detailed procedure of Condition 2 in Appendix B of the Supplementary Information.

**Condition 3:** single-AI creation. In this condition, we explored the creative process and outcomes when a single AI agent generated ideas independently, without collaboration or iterative refinement. The AI agent in this condition was set to a temperature of 0.5. For each task, the AI agent was prompted with the task description and instructed to generate one creative solution of 50-100 words. The agent was asked to provide the solution it considered the most creative. This single-round, single-agent ideation process allowed us to assess the AI agent's baseline creative performance without the influence of collaboration or iterative refinement.

**Condition 4:** human-human co-creation. Participants were recruited through the Behavioural Lab participant pool at a UK university in December 2023. The study was advertised as a 40-minute paired collaboration study for £7, and the study sessions were conducted online via Zoom. In total, 211 participants signed up and showed up during the Zoom sessions. We randomly assigned participants in each session into pairs, and for those participants who were not paired due to uneven numbers of attendance, they were assigned to collaborate with an AI chatbot, and their data was deleted from the sample for this condition. In the end, we collected data from 204 participants (102 pairs; $Mean_{Age}$=28.45, $SD_{Age}$=10.63, 71% female, 72% at least a bachelor's degree). Each pair was tasked with collaborating on the same three creative tasks used in the AI conditions. We randomized the order of the three tasks to control for potential order effects. The participants interacted with each other through an online chatroom interface created using SMARTRIQS, which allowed for natural, real-time interaction and random matching between participants[10]. They were instructed to work together and generate creative solutions for each task, with a target length of 50-100 words per solution. The pairs were given the freedom to collaborate in any way they deemed effective, such as brainstorming ideas, providing feedback to each other, and iteratively refining their solutions. After completing the creative tasks, participants were asked to complete a brief survey to provide subjective ratings of their collaboration experience and demographic information. On average, 6.11 interactions were taken between the two AI agents for task 1, 6.68 interactions for task 2, and 7.52 interactions for task 3.

Overall, the study compiled a total of 404 solutions across the four conditions ($N_{condition1}$ = 100, $N_{condition2}$ = 100, $N_{condition3}$ = 102, $N_{condition4}$ = 102). This experimental design was replicated across three different tasks, resulting in a total of 1,212 solutions.

**Measures**

We measured the creativity, novelty, and usefulness of the collected solutions using the consensual assessment technique[11,12]. We recruited and trained three evaluators to independently evaluate the creativity, novelty, and usefulness of these solutions. After their initial evaluation, we performed an inter-rater reliability test, and evaluators were called together later to reach a consensus on extremely disagreed solutions. The solutions of the highest and lowest scores in creativity, novelty, and usefulness for each task are included in Appendices C, D, and E of the Supplementary Information.